\documentclass[aps,prl,twocolumn,superscriptaddress,floatfix]{revtex4-1} 

\usepackage{amssymb}
\usepackage{amsmath}
\usepackage{amscd}
\usepackage{graphicx}
\usepackage{amsbsy}
\usepackage{color}

\graphicspath{{./FIGS/}}

\begin{document}

\author{Flavio Romano}
\affiliation{{Dipartimento di Fisica, Universit\`a di Roma {\em La Sapienza}, Piazzale A. Moro 2, 00185 Roma, Italy}}
\email{flavio.romano@gmail.com}
\author{Francesco Sciortino}
\affiliation{{Dipartimento di Fisica and CNR--ISC, Universit\`a di Roma {\em La Sapienza}, Piazzale A. Moro 2, 00185 Roma, Italy}}
\email{francesco.sciortino@uniroma1.it}
\title{Crystal phases of two dimensional assembly of triblock Janus particles}

\date{\today}

\begin{abstract}
Recent experimental work on spherical colloidal particles decorated with
two hydrophobic poles separated by an electrically--charged middle band
(triblock Janus particles) has documented self--assembly into a Kagome
two--dimensional lattice, when particles are confined by gravity at the
bottom of the sample holder [Q. Chen {\em et. al.}, Nature, in press]. Here
we assess the ability of a previously proposed simple two--patch effective
potential to reproduce the experimental findings. We show that the
effective potential is able to reproduce the observed crystallization
pathway in the Kagome structure. Based on free energy calculations, we also
show that the Kagome lattice is stable at low temperatures and low
pressure, but that it transforms into a hexagonal lattice with alternating
attractive and repulsive bands on increasing pressure.
\end{abstract}


\maketitle

Chemical or physical patterning the surface of particles provides an
effective way of modulating the interaction between colloidal particles.
The possibility of designing particles that interact via a non--spherical
potential opens up a wealth of new possibilities as envisioned in the
anisotropy axis space by Glotzer and Solomon~\cite{Glo07a}. The challenge
faced by physicists, chemical engineers and material scientists is thus to
organize these new geometries into structures for functional materials and
devices via self--assembly, the spontaneous organization of matter into
desired arrangements. The aim is to achieve --- via the rational design of
elementary building blocks (i.e. the particles) --- pre--defined specific,
ordered or disordered, structures. Research in this direction is very
active~\cite{Paw10a}, even though most of the experimental efforts are
still in the direction of acquiring control over the desired distribution
of patch widths and number~\cite{Manoh_03, mohwald, kegel}, more than on
the collective behavior of the particle themselves (with noteworthy
exceptions~\cite{Glo06a, oldgranick1, oldgranick2, gangnature1}).
Self--assembly of patchy particles has been the focus of a large number of
theoretical and numerical investigations~\cite{Glo07a, bianchiprl, Hui08a,
Cac06a, doyepccp, Rom09a, Rom10a, doye1, doye2}, which have revealed a
wealth of novel physical phenomena, some of which having an analogous
counterpart in atomic or molecular systems~\cite{fsepjb, russoprl}.

\begin{figure*}[tb]
\centering
\includegraphics[width=4cm]{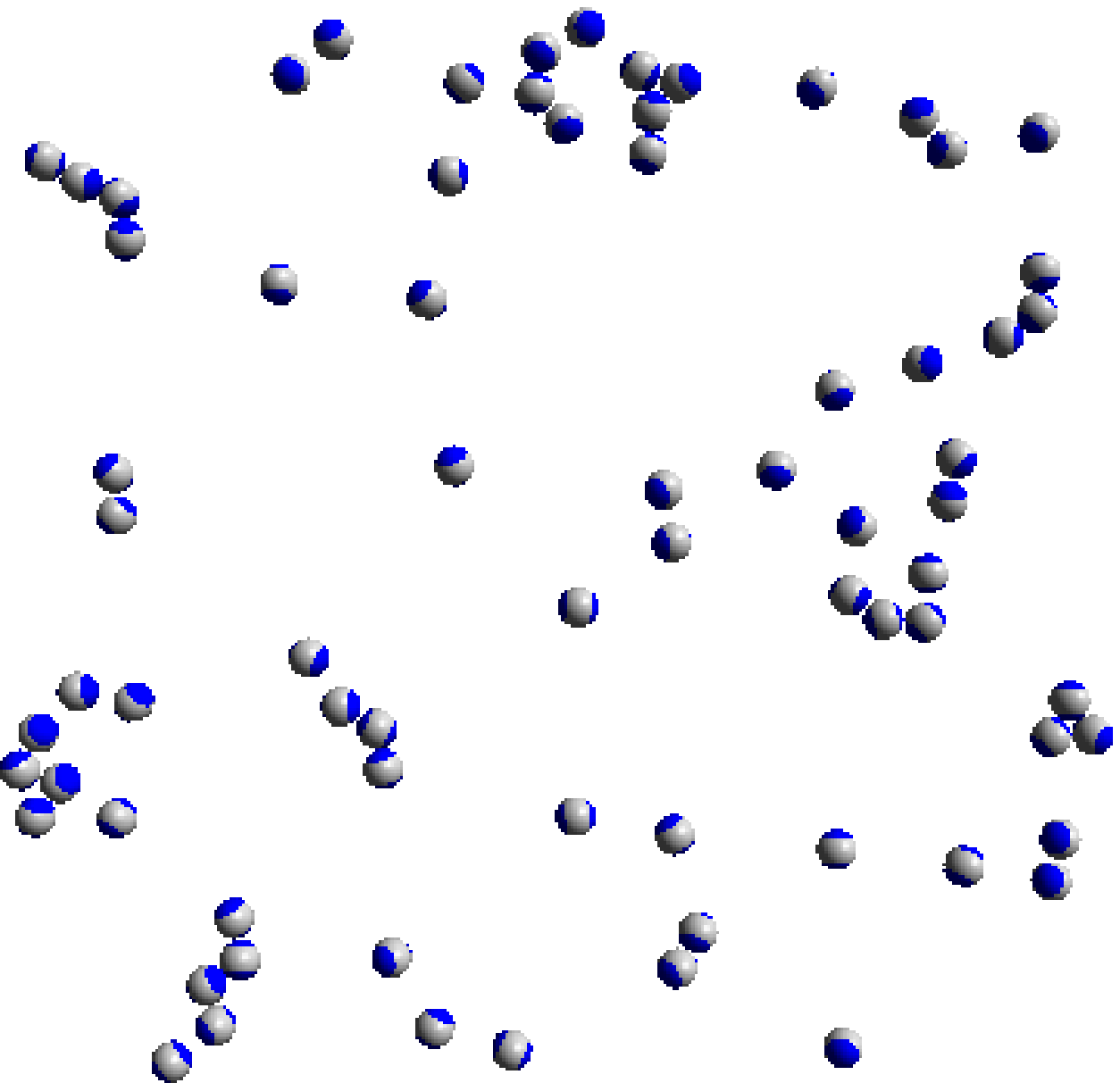}
\includegraphics[width=4cm]{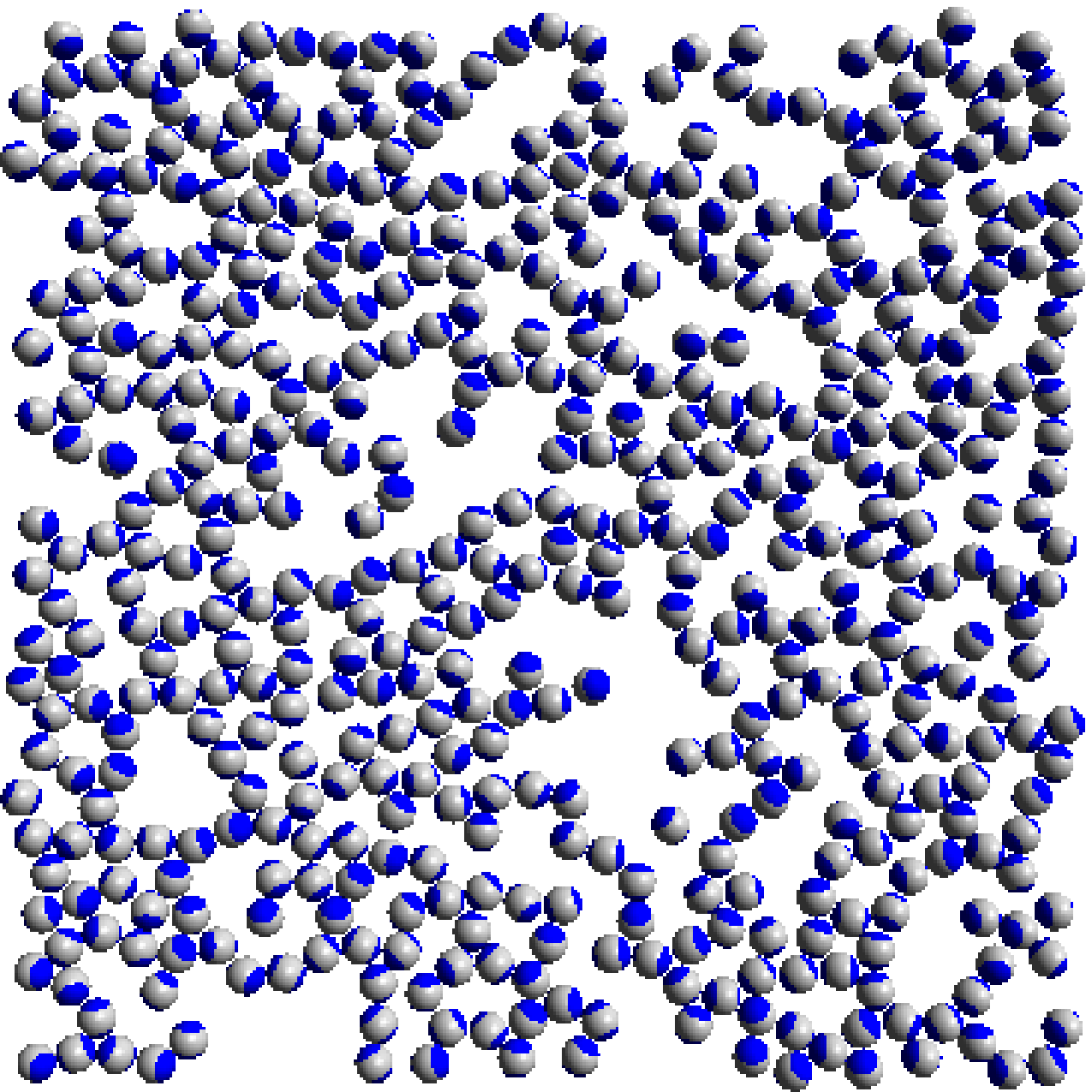}
\includegraphics[width=4cm]{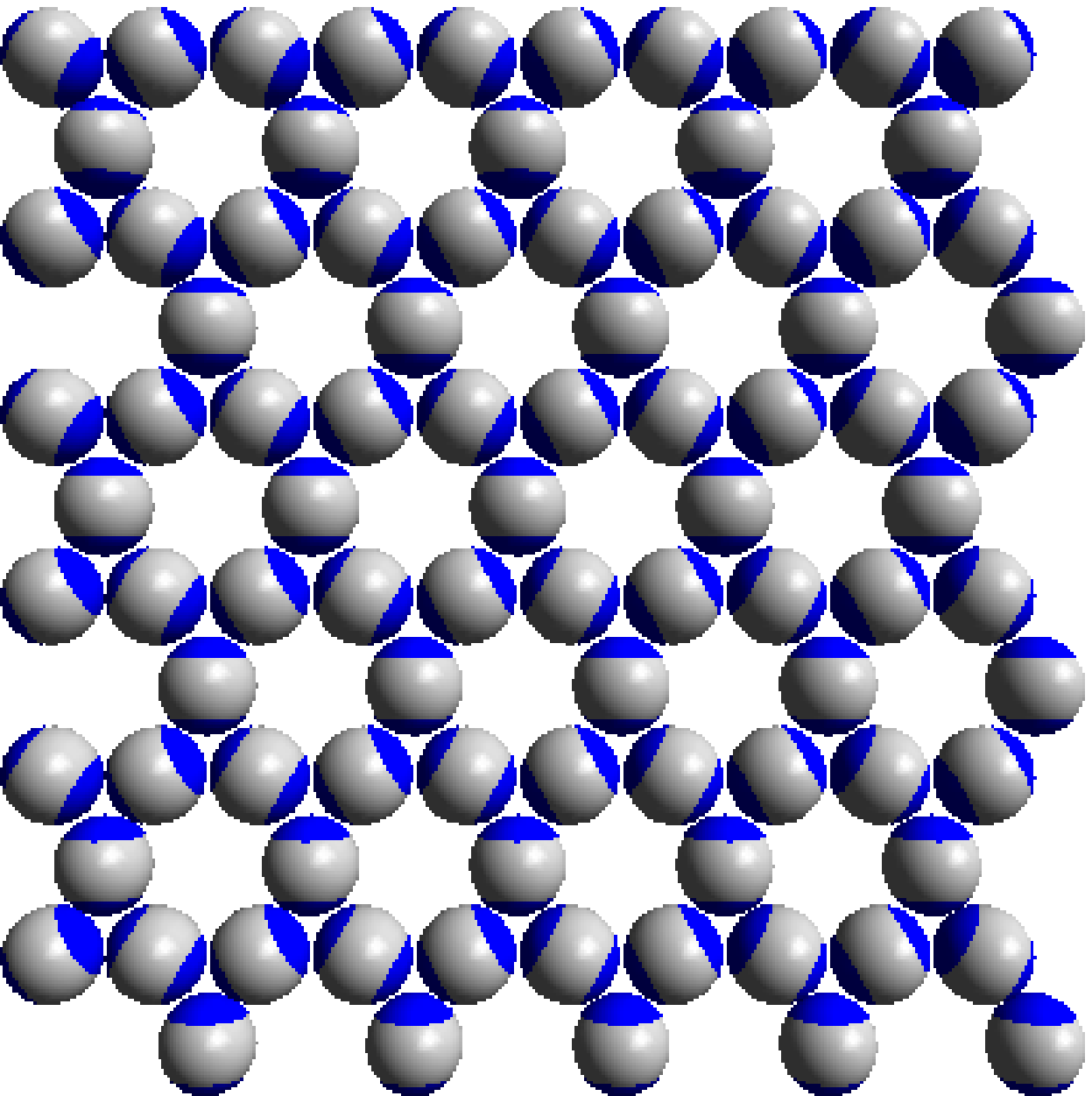}
\includegraphics[width=4cm]{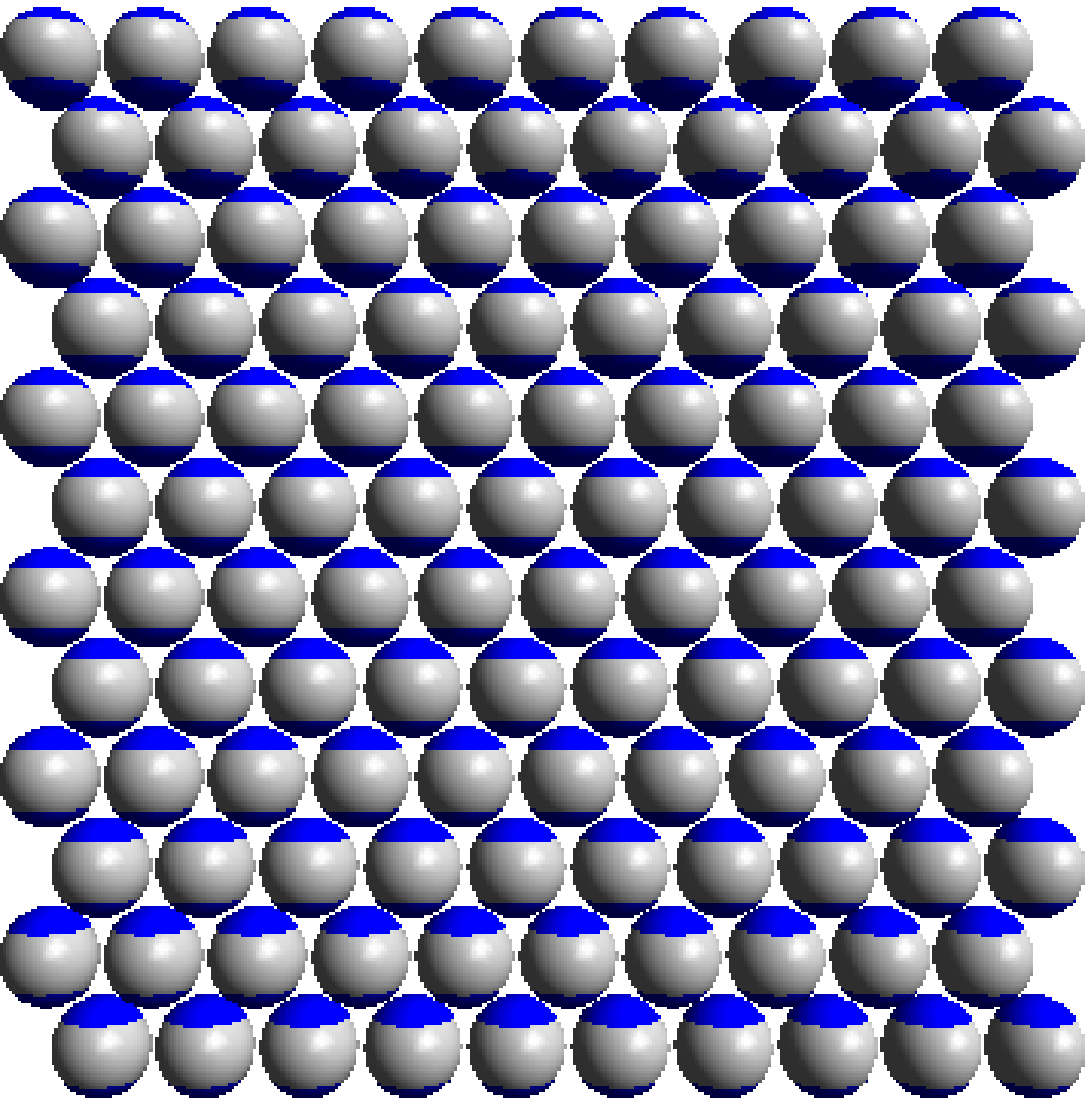}
\caption{\label{fig:kagome}
From left to right: snapshot of a gas, liquid, Kagome lattice and hexagonal
lattice. The Kagome and the hexagonal crystals are formed respectively at
low and high pressures. Here the patch width is $\cos(\theta)=0.524$.
Patches are colored in blue, the hard--core remaining particle surface is
colored in gray. Particles are free to rotate in three dimensions but are
constrained to move on a flat surface.
}
\end{figure*}

A very recent experimental work based on spherical colloidal particles
decorated with two hydrophobic poles of tunable area, separated by an
electrically charged middle band (triblock Janus)~\cite{granicknature}
provides an excellent example of accurate synthesis of two--patch
particles accompanied by a study of the self--assembly of these particles
into an ordered structure when deposited on a flat surface. Interestingly,
the electric charge of the particles allows for a controlled switch of the
interaction via addition of salt, which effectively screens the overall
repulsion, offering the possibility to the hydrophobic attraction between
patches to express itself. After the addition of the salt, particles organize
themselves into a Kagome lattice. The crystallization kinetics can be followed
in real space in full details. The width of the patches, of the
order of 65 degrees, allows for simultaneous bonding of two particles per
patch, stabilizing the locally four--coordinated structure of the Kagome
lattice (see Fig.~\ref{fig:kagome}). Experiments also show that, when more than one layer of particles sediment,
stacked Kagome planes form. As discussed in Ref.~\cite{granicknature}, such
alternating Kagome planes could have potential application as selective
filters, where selection is potentially controlled by the two different
typical sizes of the basic Kagome structure (the triangle and the hexagon)
as well as by the different chemical characters of the two holes
(hydrophilic and hydrophobic).

The three--dimensional collective behavior of two--patch particles, modeled
via the Kern--Frenkel potential~\cite{Ker03a}, has been studied
recently~\cite{giacometti2} as a function of the patch width, interpolating
between the isotropic case, when each patch covers one hemisphere, and the
case where each of the two opposite patches can be involved only in one
bond, generating a polydisperse distribution of colloidal chains
(equilibrium polymers). In the range of angular patch width compatible with
only two bonds per patch (the same as the experimental system alluded
before), the spontaneous formation of an ordered phase in a fully bonded
closely--packed hexagonal lattice with alternating attractive bands (see
Fig.~\ref{fig:kagome}) was reported. While gravitational effects do not
allow a close comparison between experiments and simulations in three
dimensions, a theoretical study of the phase diagram of the same model in
two dimensions --- but retaining the full three--dimensional orientational
properties --- can provide a valuable test for validating the effective
potential as well as for estimating the relative stability field of the
possible crystals and of the fluid phase.

In this Letter we study the phase diagram of the two--patch Kern--Frenkel
model in two dimensions, for two different values of the patch width, both
within the angular width interval that allows for at most two bonds per
patch. In full agreement with the experiments, we observe, at comparable
pressure and interaction strength, the spontaneous nucleation of Kagome
lattice. At larger pressure, spontaneous crystal formation in the dense
hexagonal structure is observed. Such easiness to crystallize suggests that
in this system crystallization barriers are comparable to the thermal
energy at all pressures. Interestingly enough, we find that for this model
a (metastable) gas--liquid phase separation can be observed for large patch
width.

We study the Kern--Frenkel~\cite{Ker03a} two--patch model where two
attractive patches are symmetrically arranged as polar caps on a hard
sphere of diameter $\sigma$. Each patch can thus be envisioned as the
intersection of the sphere surface with a a cone of semi--amplitude $\theta$
and vertex at the center of the sphere. The model assumes that a bond (with
interaction energy $-u_0$) is established between two particles when their
center--to--center distance is less than $\sigma(1+\delta)$ and the line
connecting their centers crosses two arbitrary patches on distinct
particles. Reduced units will be used throughout this work, with $k_{\rm B}=1$, temperature $T$ in units of $u_0/k_{\rm B}$, pressure $P$ in units
of $\sigma^{-2}u_0$ and number density $\rho$ in units of
$\sigma^{-2}$. A precise definition of the model can be found in
Ref.~\cite{Ker03a} and in the Supplemetary Informations. The model has
been extensively investigated in simulation and theoretical studies of
patchy particles~\cite{Ker03a, foffi, giacometti2, Giacometti09b, Liu07a},
including integral equation~\cite{giacometti2, Giacometti09b} as well as
perturbation theories~\cite{gogelein} for anisotropic potentials. Here we focus on
the short-range value $\delta=0.05$ (comparable to the experimental value)
and two values of the patch width: the experimental values $cos(\theta=0.84)$ and the
largest value consistent with the two-bond-per-patch condition, $cos(\theta=0.524)$.

The phase diagrams have been calculated following the
methodologies reviewed in Ref.~\cite{vega-review}. Specifically, we have
calculated the free energy of crystal phases via thermodynamic integration
from the Einstein crystal~\cite{frenkel-ladd}, and the free energy of fluid
phase via thermodynamic integration from the ideal gas. With these
techniques, precise estimates of the chemical potential of the crystal and
of the fluid phase can be obtained. The Gibbs--Duhem
integration~\cite{Kofke} is then implemented to compute the coexistence
line in the pressure-temperature $P-T$ plane. Consistency checks based on
direct coexistence~\cite{Ladd77a} have been performed to validate the
results of our free energy calculations. The gas--liquid critical point and
coexistence densities have been calculated implementing the successive umbrella
sampling method~\cite{sus}. We remand to the Supplementary Informations
for technical details regarding free energy calculations.

We also perform NPT and NVT Monte Carlo (MC) simulations of a system of
1000 particles for several values of $P$ and $T$ to gather structural
information and to study spontaneous crystallization.
MC simulation runs were carried out for at least $10^{6}$ MC cycles, a
third of which were used for equilibration. The translational, rotational
and volume trial displacements were calibrated during equilibration to have
a global acceptance ratio of 0.5, 0.5 and 0.25 respectively.

\begin{figure*}[tb]
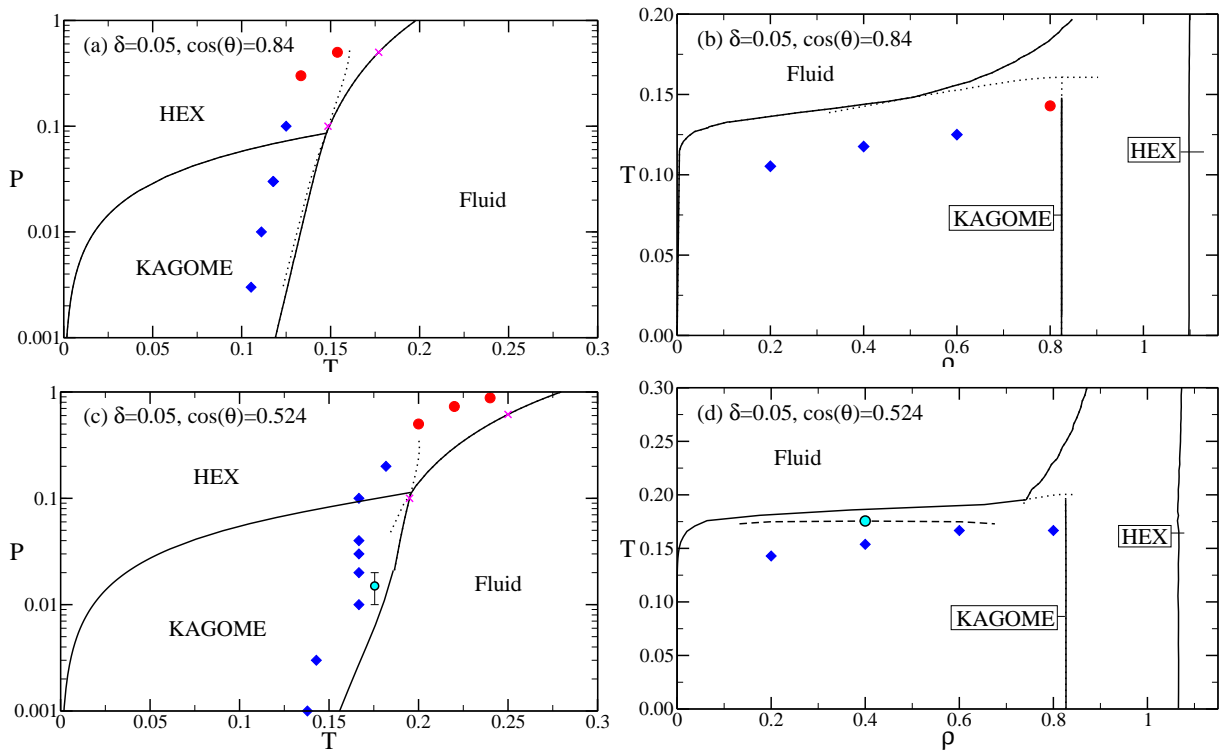

\includegraphics[width=8cm]{p-t-84}
\includegraphics[width=8cm]{t-rho-84}\\
\includegraphics[width=8cm]{p-t-524}
\includegraphics[width=8cm]{t-rho-524}\\
\caption{\label{fig:phase}
Phase diagrams in the $P-T$ representation (left column) and $T-\rho$
representation (right column) for the narrow (top row) and wide (bottom row)
patch model. Boundaries between stable phases are drawn in solid black, while
metastable phase boundaries are dotted. The cyan point in panels (c) and (d)
indicate the (metastable) gas--liquid critical point. The dashed line in panel
(d) represents the metastable gas--liquid phase separation. Blue diamonds and
red circles indicate the highest temperature at which spontaneous
crystallization into the Kagome and hexagonal lattice respectively was detected
at the corresponding $P$ or density. Crosses indicate the coexistence points
checked via direct coexistence simulations.
}
\end{figure*}

Fig.~\ref{fig:phase} shows the phase diagram for the two investigated
values of $\cos(\theta)$. In the investigated range of $T$ and $P$, two
fully bonded crystals can be predicted to possibly form: the Kagome and the
hexagonal lattices (see Fig.~\ref{fig:kagome}). The open Kagome structure
is stable at low $P$, while the dense hexagonal one at larger $P$ values.
At large $T$ a fluid phase is stable. We note that on increasing $P$ beyond
the range reported in Fig.~\ref{fig:phase} , other crystal phases appear,
including a plastic hexagonal phase and a dense polymeric phase. The
topology of the phase diagram is essentially identical, with the
coexistence lines shifted to higher $P$ and higher $T$ for the larger
width, as expected on the basis of the larger bonding volume (which
correspondingly reflects a larger virial coefficient). In agreement with
the experimental system, the Kagome structure becomes stable when the
patch--patch interaction strength becomes five to ten times the thermal
energy, a range of values at which bonds can still be thermally broken
providing an effective way to escape kinetic traps and accurately sample
the phase space.

\begin{figure}[tb]
\centering
\includegraphics[width=8cm]{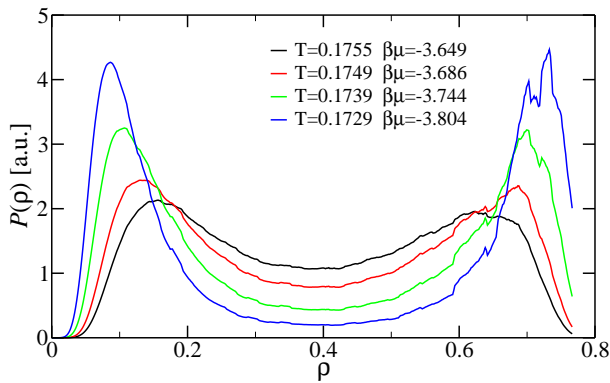}
\caption{\label{fig:pn}
Probability of the density fluctuations ${\mathcal P}(\rho)$ close to the metastable gas--liquid
critical point for the wide--patch model ($\cos(\theta)=0.524$). The
formation of two different liquid--like and gas--like states in coexistence
is clearly seen. Please note that for $\rho>0.8$ a third peak (not shown),
corresponding to the crystal, develops. For $T<0.1729$ crystallization is
so effective that it is impossible to gather enough statistics to evaluate ${\mathcal P}(\rho)$.
}
\end{figure}

To test if the fluid phase undergoes a gas--liquid phase separation at low
$T$, we investigated the behavior of the density fluctuations.
Fig.~\ref{fig:pn} shows that the distribution of the density in the grand
canonical (constant volume, $T$ and chemical potential $\mu$)
ensemble~\cite{sus} acquires the typical bimodal shape characteristic of
coexisting gas-like and liquid-like regions which characterize the system
close to criticality. Typical gas configurations are characterized by
isolated clusters, while liquid-like configurations show a percolating
network of bonds, with a large fraction of loops of different size (see
Fig.~\ref{fig:kagome}). The critical point is slightly metastable (see
Fig.~\ref{fig:phase}). As noticed in Ref.~\cite{giacometti2}, the
progressive restriction of the bonding angle plays a role analogous to the
reduction of the range in spherically interacting attractive
colloids~\cite{Vli00a}. Indeed, critical fluctuations can only be observed
for the large angular width model. In the case of the small angular width
($\cos(\theta)=0.84$), the possibility of observing the phase separation is
preempted by crystallization. In this small width case, crystal formation
is so effective that there is no time for establishing a time--independent
metastable liquid state and properly evaluate the density fluctuations. A
proper choice of the patch width and interaction range in the experimental
system, in which particles can be optically tracked, could thus provide a
way of directly observing critical fluctuations in a two--dimensional
system in real space.

Fig.~\ref{fig:phase} also shows the points where spontaneous crystal
formation in one of the two crystal forms is observed during a
constant--$NPT$ simulation. Interestingly, around $P \approx 0.1$, the
Kagome crystal develops (as a metastable form) in the region of stability
of the denser crystal, consistently with the Ostwald's rule~\cite{Ostwald}
which states that in general it is the least stable polymorph that
crystallizes first. In the present case the difference in chemical
potential between the hexagonal lattice and the Kagome lattice is only of
the order of 0.03$k_{\rm B}T$ at $P \approx 0.1$ and hence can not really
control the preferential crystallization in the less stable lattice
structure. We suggest that the preferential formation of the Kagome
structure at $P=0.1$ is the result of the fluid having a density closer to
that of the Kagome lattice than that of the denser lattice.

More relevant for colloidal applications is the phase diagram in the
$T-\rho$ plane. The Kagome structure is stable in a small density
window. For the studied model, the Kagome structure can exist only for
$(\sqrt{3}/2)/(1+\delta)^2 < \rho \sigma^2 < \sqrt{3}/2$. In the last
inequality, the upper value is controlled by excluded volume, while the
lower figure is controlled by the range of the interaction potential.

\begin{figure}[tb]
\centering
\includegraphics[width=8cm]{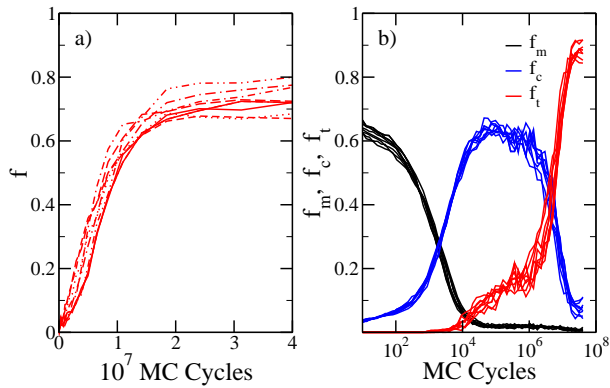}
\caption{\label{fig:crys}
Analysis of the crystallization process of the Kagome lattice. In (a) the
fraction $f$ of solid--like particles (identified as explained in
Supplementary Informations) is shown as a function of the number of MC
cycles. The number of particles in a Kagome lattice exponentially
approaches a plateau. In (b) the relative abundance of monomers, chainly
bonded and triangularly bonded particles ($f_m$, $f_c$ and $f_t$
respectively) is shown versus simulation time. The crystallization of the
Kagome lattice is preceded by the formation of extensive chain--like
bonding in the system. Both (a) and (b) show a striking resemblance with
the experimental system. In both panels the results of nine independent
$NVT$ runs are reported. In all the runs, $\delta=0.05$,
$\cos(\theta)=0.84$, $N=1000$, $\rho=0.6$ and $T=0.125$. Details on how we
identified particle types are given in Supplementary Informations.
}
\end{figure}

The crystallization kinetics of triblock Janus particles has also been
investigated experimentally, by optical microscopy, providing a detailed
description of the formation of the Kagome lattice. To provide evidence
that the Kern-Frenkel potential is able not only to reproduce the
thermodynamic properties, but also the crystallization pathway, we report
in Fig.~\ref{fig:crys} the time evolution of the fraction of crystalline
particles as well as the time dependence of the concentration of particular
geometrical arrangements of the particles. Specifically, we focus
(following Ref.~\cite{granicknature}) on particles in chains, particles
forming triangular bond loops (the unit element of the Kagome lattice), and
un-bonded particles. As time grows, the number of un-bonded isolated
particles decreases in favor of the formation of chains of oppositely
bonded particles which then restructure themselves to form the triangular
elements of the ordered lattice. A movie of the crystallization process is
available in the Supplementary Informations. Comparing Fig.~\ref{fig:crys} with
Fig.~2 and 3 of Ref.~\cite{granicknature} one sees that indeed the
effective potential properly describes the kinetics of crystal formation.


To summarize, particles made by a repulsive core and attractive patches are
clearly one of the most promising model systems for generating, via a
rational design~\cite{granicknature}, specific structures. The versatility
of the method has been already proved for one and two-patch particles,
revealing in both cases interesting assembly processes. In the case of one
patch, particles aggregates in micelle or in branched linear
clusters~\cite{oldgranick1,oldgranick2}. Interestingly, also in the
one-patch case, the simple Kern--Frenkel potential has been shown to
properly reproduce the experimentally observed structures~\cite{janusprl}.
The ability of accurately describing Janus triblock particles with the same
model, by only changing the geometry of the patches as in the experimental
system, is particularly rewarding and provides a strong support for the use
of such model for predicting the self--assembly properties of this class of
patchy colloids. The possibility of numerically exploring the sensitivity
of the phase diagram to the parameters (patch width and interaction range)
entering in the interaction potential provides an important instrument and
a guide to the design of these new particles to obtain specif structures by
self--assembly.

It is interesting to observe the analogies between the two--dimensional
phase diagram of these two--patch two--bonds particles with the
three--dimensional phase diagram of tetrahedral particles in which each
patch can be only engaged in one bond~\cite{Rom09a,Rom10a}. In both cases,
the phase diagram is characterized by the competition between an open
(diamond in 3D, Kagome in 2D) crystal and a denser one (BCC in 3D,
hexagonal in 2D). In both cases, smaller angular patches favors crystal
formation, completely pre--empting the possibility of forming a metastable
liquid state. In both cases, the propensity for crystallization arises from
a significant difference in the slope of $\beta \mu$ vs $T$ at the melting
temperature.

In this contribution we have focused on the two--dimensional phase diagram,
showing that the open Kagome structure is stable at low $P$ and
$T$. An accurate study based on the comparison of the free energy
of the different crystal forms of the corresponding three--dimensional
system has not been performed so far. Hence, it is not clear if the Kagome
structure would still self--assemble even if the system were not confined by
gravity on a surface. After the validation of the model reported here,
theoretical evaluation can help answering this important question.

\section{Acknolwdgements}
We thank S. Granick for sending us a pre--print of
Ref.~\cite{granicknature} prior to publication and C. De Michele for
providing us with the program with which the graphic representations were
made. We acknowledge support from ERC-226207-PATCHYCOLLOIDS and ITN-234810-COMPLOIDS.

\bibliography{kagome}

\end{document}